\documentclass[journal,transmag]{IEEEtran}
\IEEEoverridecommandlockouts

\usepackage[cmex10]{amsmath}
\usepackage[ruled, vlined, linesnumbered]{algorithm2e}
\usepackage{graphicx,adjustbox}
\usepackage[xindy,acronym,nonumberlist]{glossaries}
\usepackage{arydshln}
\usepackage{tabularx}
\usepackage{booktabs}
\usepackage{makecell}
\usepackage{amssymb}
\usepackage{colortbl}
\usepackage{xcolor}
\usepackage{longtable}

\definecolor{grey}{rgb}{0.8,0.8,0.8}
\definecolor{lightgrey}{rgb}{0.9,0.9,0.9}
\usepackage{cite}

\usepackage{caption}
\usepackage{subcaption}

\usepackage{hyperref}
\hypersetup{colorlinks,urlcolor=blue, citecolor=blue, linkcolor=blue}

\usepackage{lipsum}
\usepackage[disable]{todonotes}

\usepackage{amsmath,tikz, tikz-3dplot, pgfplots}
\usepackage{pgfplots}
\usepackage{pgfplotstable}
\usetikzlibrary{matrix,decorations.pathreplacing}
\usetikzlibrary{shapes,arrows, patterns}

\newtheorem{prop}{Proposition}

\newacronym{RFEC}{RFEC}{Remote Field Eddy Current}
\newacronym{PEC}{PEC}{Pulse Eddy Current}
\newacronym{LLS}{LLS}{Linear Least Squares}
\newacronym{RFT}{RFT}{Remote-Field Technologies}
\newacronym{MI}{MI}{Mutual Information}
\newacronym{PDF}{PDF}{Probability Density Function}
\newacronym{NDE}{NDE}{Non-Destructive Evaluation}
\newacronym{GP}{GP}{Gaussian Process}
\newacronym{NDT}{NDT}{Non-Destructive Testing}
\newacronym{SVM}{SVM}{Support Vector Machine}
\newacronym{FEA}{FEA}{Finite Element Analysis}
\newacronym{ROR}{ROR}{Radius Outliers Removal}
\newacronym{SOR}{SOR}{Statistical Outliers Removal}
\newacronym{BandS}{B\&S}{Bell and Spigot}
\newacronym{SLAM}{SLAM}{Simultaneous Localisation and Mapping}
\newacronym{MFL}{MFL}{Magnetic Flux Leakage}
\newacronym{BEM}{BEM}{Eddy Current Broadband Electromagnetic}
\newacronym{SQUID}{SQUID}{Superconducting QUantum Interference Device}
\newacronym{LASSO}{LASSO}{Least Absolute Shrinkage and Selection Operation}
\newacronym{MSE}{MSE}{Mean Square Error}
\newacronym{TGP}{TGP}{Twin Gaussian Process}
\newacronym{1D}{1D}{One Dimensional}
\newacronym{2D}{2D}{Two Dimensional}
\newacronym{3D}{3D}{Three Dimensional}
\newacronym{GMR}{GMR}{Giant Magneto Resistance}

\newacronym{RIMLS}{RIMLS}{Robust Implicit Moving Least Squares}

\begin{document}
\title{Remote Field Eddy Current Signal Deconvolution and Towards Inverse Modeling}

\author{\IEEEauthorblockN{Raphael Falque\IEEEauthorrefmark{1},
Teresa Vidal-Calleja\IEEEauthorrefmark{2}, and
Jaime Valls Miro\IEEEauthorrefmark{3}}
\IEEEauthorblockA{\IEEEauthorrefmark{1,2,3}Faculty of Engineering and Information Technology, 
University of Technology Sydney, Australia}
\thanks{Manuscript submitted \today.
Corresponding author: R. Falque (email: Raphael.H.Guenot-Falque@student.uts.edu.au).}}
	
\IEEEtitleabstractindextext{%
\begin{abstract}
\glsresetall
Being able to quantify the corrosion of ferromagnetic water pipeline is a critical task for avoiding pipe failures. This task is often performed with \gls{NDE} technologies and solving the inverse modeling of the sensor is considered as the holy grail of most \gls{NDE} studies. In this paper, we propose a method that partly solves the inverse problem for the \gls{RFEC} technology. We consider \gls{RFEC} tools with two different design, the first one is the classic design with a single exciter coil and a single receiver coil, and the second design considers the replacement of the receiver coil by an array of receivers distributed along the circumference. The proposed algorithm is tested on both simulated and real dataset.
\end{abstract}

}

\maketitle

\section{Introduction}
\label{sec:introduction}
\glsresetall

Water-pipelines made of ferromagnetic materials are subject to corrosion, which can lead to expensive and dangerous pipe failures. These failures have to be predicted by assessing the quality of the water pipelines and the infrastructure has to be replaced in case of failure likely to happen. To assess the quality of the pipelines, the water industry relies on \gls{NDE} approaches which allow for a cost-effective assessment of pipelines. The \gls{RFEC} technology is one of the electromagnetic \gls{NDE} technology allowing the assessment of ferromagnetic pipes. This technology uses an inline tool that travels inside a pipe while gathering information related to the remaining non-corroded thickness of the pipe wall. The study of the data gathered by the tool allowed to plan replacement of the corroded pipes section and ultimately avoiding the pipe failures.

\begin{figure}
	\centering
	\includegraphics[width=1\linewidth]{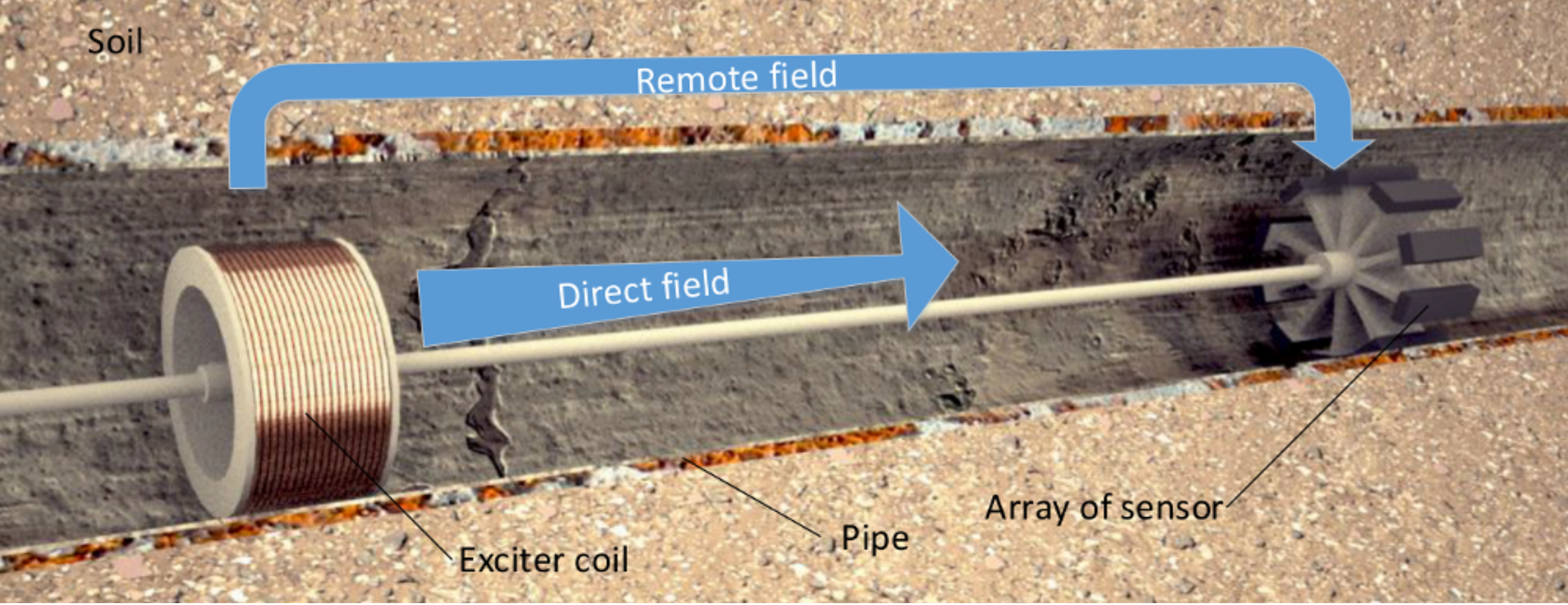}
	\caption{Schematic of the RFEC technology. The electromagnetic field propagates following the direct and the remote direction. Due to the eddy currents, the direct field  is quickly attenuated resulting in the remote field dominating the direct field is the remote area.}
	\label{fig:RFEC_schematic}
\end{figure}

\gls{RFEC} tools were originally designed with an exciter coils and a receiver coil located at a remote distance along the axial direction~\cite{MacLean1951}. It can be shown through \gls{FEA} that the electromagnetic field generated by the exciter coil flows outward the pipe near the exciter coil and flows inward the pipe at a remote location, hence the placement of the receiver coil in this remote area~\cite{Lord1988}. This phenomenon, referred as \textit{double through wall penetration} in the literature, is summarized in Figure~\ref{fig:RFEC_schematic}. In practice, the electromagnetic field flows in every direction, and the measurement in the remote area is the result of both the direct field and the remote field; however, the direct field is strongly attenuated due to eddy currents and can be neglected compared to the remote field in the remote area. Each time the electromagnetic field crosses the pipe wall, its amplitude is attenuated and, its phase is delayed proportionally to the non-corroded thickness of the pipe wall~\cite{Palanisamy1987}, making this technology particularly interesting for \gls{NDE}. As a result, the sensor measurements of an \gls{RFEC} tool are usually both the amplitude of the electromagnetic field measured by the receiver coil and the phase-shift between the generated and measured electromagnetic field~\cite{Sun1996, Atherton1995a}.

Most \gls{NDE} technologies based on electromagnetics have a relatively direct relationship between the sensor measurements and the non-corroded thickness of the inspected pipes. For instance, the thickness can be inferred from \gls{PEC} and \gls{MFL}\cite{Wijerathna2013} measurements using a \gls{GP}. Conversely, due to the double through wall penetration, the sensor measurements from an \gls{RFEC} tool is related to the thickness of the pipe in different areas. This spatial dependency can be shown in simple experiments such as inspecting a perfect pipe with a single defect; the sensor measurements for such experiment would see a change in signal strength when the defect is near the exciter coil and when the defect is near the receiver coil~\cite{Davoust2006}.

For simple scenarios (e.g. perfect pipe with single defect), this property can be used to have a redundant description of a single defect and therefore improve the prediction robustness. However, in the case of corroded pipelines, this ideal scenario is far from the reality; thus, the signal has to be deconvolved. 

For tools designed axisymmetrically with a single exciter coil and a single exciter coil, it has been shown that the signal deconvolution and the inverse problem can be solved with a \gls{LLS} formulation~\cite{Falque2016}. However, this formulation requires several measurements to infer a single thickness. This is an inherent property of \gls{LLS} which requires more equations than the number of inferred variables to have stable computations. This problem can be solved by adding a second receiver coil and using a Wiener deconvolution filter~\cite{Luo2016}. While such design is commonly used in simulated environments, the additional coil leads to a bulky design which is not convenient for practical implementations.

In this article, we consider the signal deconvolution for different tool design and discuss the step required to solve the inverse model of the \gls{RFEC} technology. More precisely, we propose a novel formulation of the signal deconvolution in Section~\ref{sec:1D_deconvolution}, some insights about the behavior of the magnetic field in a \gls{3D} environment are discussed in Section~\ref{sec:mf_behavior}, using these insights we proposed a signal deconvolution for a tool with a circumferential array of receivers in Section~\ref{sec:2D_deconvolution}. In Section~\ref{sec:discussion}, we discuss the steps towards the inverse model.

\section*{Nomenclature}
\begin{tabular}{cl}
	$B$				    & amplitude of the magnetic field \\
	$\varphi$			& phase-shift between exciter coil and receiver\\
	$\omega$            & circular frequency \\
	$\mu$               & magnetic permeability \\
	$\sigma$            & electrical conductivity \\
	$\boldsymbol{y}$    & 1D sensor measurements (one receiver coil) \\
	$\boldsymbol{Y}$    & 2D sensor measurements (array of receiver) \\
	$t_i$               & local pipe thickness measurement\\
	$\boldsymbol{t}$    & thickness profile \\
	$w_i$               & parameter learned from the direct model \\
	$\boldsymbol{w}$    & set of the direct model parameters \\
\end{tabular}

\section{Method}
\label{sec:method}

\textbf{Notations:} In the article, single variables are set in lowercase, such as in $a$, arrays and matrices are defined in bold with arrays in lowercase, such as in $\boldsymbol{a}$, and matrices with uppercase, such as in $\boldsymbol{A}$. Sets are referred to with uppercase such as $A$.

\textbf{Problem formulation:} In this paper, we define the inverse modeling as the result of two different task; (i) the signal deconvolution which remove the spatial dependency of the measurement with several areas of the pipe, and (ii) the transformation from sensor space into the thickness space. More formally, if the direct modeling consists of finding the function $h$ such that $h:\boldsymbol{t}\rightarrow y$, with $\boldsymbol{t}$ a set of local thicknesses of inspected pipe and $y$ the corresponding sensor measurement. The inverse model is then defined as finding the function $h^{-1}$ such that $h^{-1}:\boldsymbol{y}\rightarrow t$. The signal convolution is then defined as finding the function $f$ which remove the spatial dependency in the sensor measurements such that $f:\boldsymbol{y}(\boldsymbol{t})\rightarrow y'(t)$, and the space transformation $h$ maps the deconvolved measurements into the pipe thickness space such as $g: y'(t)\rightarrow t$. As a result, the inverse function $h^{-1}$ is defined by $h^{-1} = f\circ g$.

\subsection{1D signal deconvolution}
\label{sec:1D_deconvolution}
\begin{figure}
	\centering
	\includegraphics[width=1\linewidth]{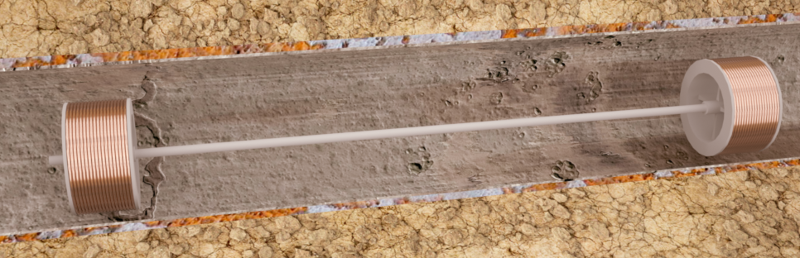}
	\caption{Sketch of an RFEC tool based on two distinctive coils. The sensor measurements are taken for different positions of the tool within the pipe and are thus defined as $y = \boldsymbol{y}(x)$, with x the position of the tool within the pipe.}
	\label{fig:RFEC_two_coil}
\end{figure}

Let us now consider the signal deconvolution for a tool made of two distinctive coils, with one of the coils used as an exciter coil and the second one used as a receiver coil. The distance between each coil is chosen in order to have the receiver coil located in the remote area. An illustration of such a tool is shown in Figure~\ref{fig:RFEC_two_coil}.

In a precedent work~\cite{Falque2016}, it has been shown that the \gls{RFEC} signal deconvolution, and under certain circumstances the inverse problem, can be formulated with an \gls{LLS}. The parametric nature of this model allows for a physical interpretation of the phenomenon. In this section, we propose a more explicit algorithm which solves the signal deconvolution. By starting from the least-square formulation established in~\cite{Falque2016} (Equation 10), and disregarding the notion of piecewise-constant pipeline profile approximation, we have the following relation:
\begin{equation}
\boldsymbol{y} = \boldsymbol{W} \boldsymbol{t} + y_0,
\label{eq:inverse_pb_1}
\end{equation}
where $\boldsymbol{W}$ is defined as
\begin{equation}
\boldsymbol{W} = 
\begin{tikzpicture}[baseline=(current bounding box.center)]
\matrix (m) [matrix of math nodes,nodes in empty cells,right delimiter={]},left delimiter={[} ]{
	w_1 & 0        &\dots        &0        & w_k     & 0         & \dots     & 0\\
	0     &          &         &          &           &           &          & \vdots \\
	\vdots     &          &         &          &          &          &          & 0 \\
	0     & \dots     & 0         &w_1     & 0        &\dots        &0        & w_k\\
} ;
\draw[loosely dotted] (m-1-1)-- (m-4-4);
\draw[loosely dotted] (m-1-2)-- (m-4-5);
\draw[loosely dotted] (m-1-4)-- (m-4-7);
\draw[loosely dotted] (m-1-5)-- (m-4-8);
\draw[loosely dotted] (m-2-1)-- (m-4-3);
\draw[loosely dotted] (m-1-6)-- (m-3-8);
\end{tikzpicture}.
\label{eq:B-less}
\end{equation}
$\boldsymbol{W}$ is a $j$ by $n$ matrix, with $n$ the number of thickness element, and $j$ is the number of sensor measurement. While we are defining the indices, $k$ is the number of measurements gathered while the tool travels its own length, i.e., the distance between the exciter coil and the array of receivers\footnote{Despite having $i = k$ numerically speaking, we differentiate the notation as they refer to different concepts.}.

Equation~\eqref{eq:B-less} can be reformulated into a more explicit formulation by splitting $\boldsymbol{W}$ into two matrices: $\boldsymbol{W}^{\text{bg}}$ and $\boldsymbol{W}^{\text{fg}}$ such as:
\begin{multline}
\boldsymbol{W} = 
\begin{tikzpicture}[baseline=(current bounding box.center)]
\matrix (m) [matrix of math nodes,nodes in empty cells,right delimiter={]},left delimiter={[}, ampersand replacement=\&]{
	w_1    \& 0         \&              \&             \&    \&        \& 0\\
	0         \&                 \&              \&              \&               \& \\
	\vdots     \&              \&             \&              \&               \& \\
	0         \& \dots        \& 0         \& w_1     \& 0    \& \dots    \& 0\\
} ;
\draw[loosely dotted] (m-1-1)-- (m-4-4);
\draw[loosely dotted] (m-1-2)-- (m-4-5);
\draw[loosely dotted] (m-2-1)-- (m-4-3);
\draw[loosely dotted] (m-1-7)-- (m-4-7);
\draw[loosely dotted] (m-1-2)-- (m-1-7);
\draw[thick, decorate,decoration={brace,amplitude=10pt,mirror},xshift=0.4pt,yshift=-0.4pt](-2.4,-1.5) -- (2.4,-1.5) node[black,midway,yshift=-0.6cm] {\footnotesize $\boldsymbol{W}^{\text{fg}}$};
\end{tikzpicture}\\
+
\begin{tikzpicture}[baseline=(current bounding box.center)]
\matrix (m) [matrix of math nodes,nodes in empty cells,right delimiter={]},left delimiter={[} , ampersand replacement=\&]{
	0   \&     \dots    \& 0            \& w_k    \& 0    \&    \dots    \& 0\\
	\&                 \&              \&              \&    \&               \& \vdots \\
	\&              \&             \&              \&    \&               \& 0 \\
	0   \&            \&              \&             \&    \& 0            \& w_k \\
} ;
\draw[loosely dotted] (m-1-1)-- (m-4-1);
\draw[loosely dotted] (m-1-3)-- (m-4-6);
\draw[loosely dotted] (m-1-4)-- (m-4-7);
\draw[loosely dotted] (m-1-5)-- (m-3-7);
\draw[loosely dotted] (m-4-1)-- (m-4-6);
\draw[thick, decorate,decoration={brace,amplitude=10pt,mirror},xshift=0.4pt,yshift=-0.4pt](-2.4,-1.5) -- (2.4,-1.5) node[black,midway,yshift=-0.6cm] {\footnotesize $\boldsymbol{W}^{\text{bg}}$};
\end{tikzpicture},
\label{eq:B_decomposition}
\end{multline}
where $\boldsymbol{W}^{\text{bg}}$ can be physically interpreted as the attenuation due to the magnetic field flowing outward from the pipe near the exciter coil and $\boldsymbol{W}^{\text{fg}}$ the attenuation due to the magnetic field flowing inward to the pipe near the receiver. Equation~\eqref{eq:inverse_pb_1} can then be reformulated as
\begin{equation}
\boldsymbol{y} = \boldsymbol{W}^{\text{fg}} \boldsymbol{t} + \boldsymbol{W}^{\text{bg}} \boldsymbol{t}  + y_0,
\label{eq:inverse_pb_2}
\end{equation}
which results into
\begin{equation}
\begin{aligned}
\boldsymbol{y} & = w_1 \boldsymbol{t}_{1:j} + w_k \boldsymbol{t}_{i:n}  + y_0,\\
& = 	w_1\boldsymbol{t}_{1:j} + 
w_k\boldsymbol{t}_{i:n} +
\dfrac{w_1}{w_1+w_k}y_0 + 
\dfrac{w_k}{w_1+w_k}y_0,\\
& =  w_1\underbrace{\bigg(\boldsymbol{t}_{1:j} + \dfrac{1}{w_1+w_k}y_0\bigg)}_{\boldsymbol{y}^{\text{fg}}} + 
w_k\underbrace{\bigg(\boldsymbol{t}_{i:n} + \dfrac{1}{w_1+w_k}y_0\bigg)}_{\boldsymbol{y}^{\text{bg}}}.
\label{eq:inverse_pb_4}
\end{aligned}
\end{equation}  
This formulation states clearly how the convolution of the signal is happening from a \gls{1D} point of view. The final sensor measurement $\textbf{y}$ can be considered as a sum of the foreground contribution $\boldsymbol{y}^{\text{fg}}$ and the background contribution $\boldsymbol{y}^{\text{bg}}$. If the thickness measurement is not needed, meaning there is no need to solve the inverse problem, the signal deconvolution can be obtained by maximizing the similarity between $\boldsymbol{y}^{\text{bg}}$ and $\boldsymbol{y}^{\text{fg}}$.

Let us consider the problem of maximizing of the similarity between $\boldsymbol{y}^{\text{bg}}$ and $\boldsymbol{y}^{\text{fg}}$ as a gradient descent optimization 
\begin{equation}
\boldsymbol{y}^{\text{bg}} = \boldsymbol{y}^{\text{bg}} - \gamma\nabla \boldsymbol{e}
\end{equation}
with the error function $\boldsymbol{e}$ defined as
\begin{equation}
\boldsymbol{e} = [\boldsymbol{y}^{\text{fg}}_{i:j} - \boldsymbol{y}^{\text{bg}}_{1:(j-i)}]^2.
\end{equation}
The gradient is then defined as follows,
\begin{equation}\label{xx}
\begin{split}
\nabla\boldsymbol{e} & = \dfrac{\partial \boldsymbol{e}}{\partial\boldsymbol{y}^{\text{bg}}}\\
& = \dfrac{\partial [\boldsymbol{y}^{\text{fg}}_{i:j} - \boldsymbol{y}^{\text{bg}}_{1:(j-i)}]^2}{\partial\boldsymbol{y}^{\text{bg}}} \\
& = 2(\boldsymbol{y}^{\text{fg}}_{i:j} - \boldsymbol{y}^{\text{bg}}_{1:(j-i)}). \\
\end{split}
\end{equation}
We then merge the constant factor 2 within $\gamma$. The final formulation of the gradient descent is then defined as
\begin{equation}
\boldsymbol{y}^{\text{bg}}_{1:j-i} = \boldsymbol{y}^{\text{bg}}_{1:j-i} - \gamma(\boldsymbol{y}^{\text{fg}}_{i:j} - \boldsymbol{y}^{\text{bg}}_{1:j-i}).
\label{eq:deconvolution}
\end{equation}
From Equation~\eqref{eq:deconvolution}, the \gls{1D} signal deconvolution can then be performed in an iterative fashion as described in Algorithm~\ref{alg:de-convolution}.

\begin{algorithm}
	\DontPrintSemicolon
	\KwIn{$\boldsymbol{y}$, $\gamma$, $\epsilon$, $w_1$, $w_k$}
	\KwOut{$\boldsymbol{y}^{\text{bg}}$, $\boldsymbol{y}^{\text{fg}}$}
	\SetKwBlock{Begin}{function}{end function}
	\Begin($\text{Deconvolution} {(} \boldsymbol{y}, \gamma, \epsilon {)}$)
	{
		$\boldsymbol{y}^{\text{fg}} \leftarrow \boldsymbol{y}$\;
		$\boldsymbol{y}^{\text{bg}} \leftarrow \boldsymbol{0}$\;
		\While{$e > \epsilon$}{
			$\boldsymbol{y}^{\text{bg}}_{1:j-i} = \boldsymbol{y}^{\text{bg}}_{1:j-i} - \gamma(\boldsymbol{y}^{\text{fg}}_{i:j} - \boldsymbol{y}^{\text{bg}}_{1:j-i})$\;
			$\boldsymbol{y}^{\text{bg}} = \text{smooth}(\boldsymbol{y}^{\text{bg}})$\;
			$\boldsymbol{y}^{\text{fg}} = \dfrac{1}{w_1}(\boldsymbol{y} - \dfrac{1}{w_k}\boldsymbol{y}^{\text{bg}})$\;
			$e = \dfrac{1}{j}\sum_{l = 1}^{j-i}|\boldsymbol{y}^{\text{fg}}_{l+i} - \boldsymbol{y}^{\text{bg}}_l|$\;
		}
		\textbf{return} $\boldsymbol{y}^{\text{bg}}$, $\boldsymbol{y}^{\text{fg}}$\;
	}
	\caption{Signal deconvolution}\label{alg:de-convolution}
\end{algorithm}

The smoothing of the signal, defined in the $6^{th}$ line of Algorithm~\ref{alg:de-convolution}, allows obtaining a more robust solution by avoiding diverging inferred measurement. It also forces the magnetic field to be continuous, which is a requirement of Gauss's law for magnetism ($\nabla\boldsymbol{B}=0$). The $7^{th}$ line is just a reformulation of Equation~\eqref{eq:inverse_pb_4}.

\subsection{Behavior of the magnetic field in the 3D space}
\label{sec:mf_behavior}
\begin{figure}
	\centering
	\includegraphics[width=1\linewidth]{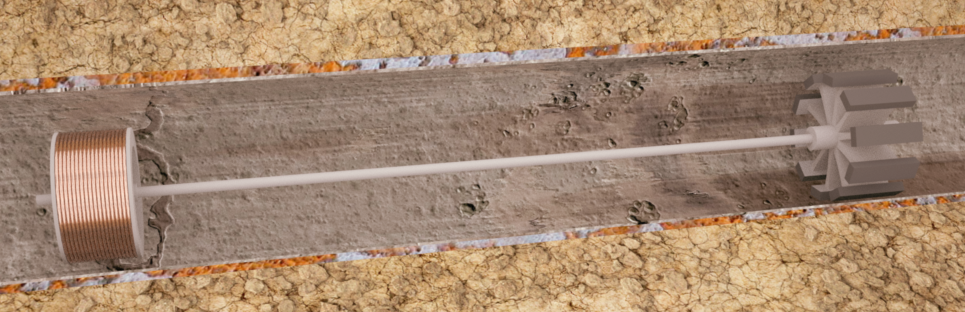}
	\caption{Sketch of an RFEC tool based on an exciter coil and an array of receiver distributed along the circumferential direction. The sensor measurements are taken for different positions of the tool within the pipe and are thus defined as $y = \boldsymbol{Y}(x, \theta)$, with x the position of the tool within the pipe and $\theta$ the position of the receiver with respect to the crown.}
	\label{fig:RFT_sensor_array}
\end{figure}

When assessing large pipelines, small defects can not be perceived with a \gls{2D} axisymmetric tool. In this situation, a tool able to acquire measurements which are more accurate than circumferential measurements is required. It can be achieved by replacing the receiver coil by an array of receiver distributed along the circumference. The distributed sensors can be chosen amongst Hall effect sensors, \gls{GMR} sensors\cite{Pasadas2013}, or coils depending on the signal to noise required. A sketch of such a tool with an array of receivers distributed along the circumference is shown in Figure~\ref{fig:RFT_sensor_array}.

To understand the difference in the measurement between the axisymmetric tool shown in Figure~\ref{fig:RFEC_two_coil} and the array of sensors shown in Figure~\ref{fig:RFT_sensor_array}, one has to understand how the magnetic field propagates from the exciter coil to the remote area. This phenomenon has been analyzed with \gls{3D} \gls{FEA} simulations in\cite{Falque2014}. We recall the results here and introduce the dataset which will be later used for validation of the \gls{3D} signal deconvolution.

\begin{figure}[h]
	\centering
	\includegraphics[width=\linewidth]{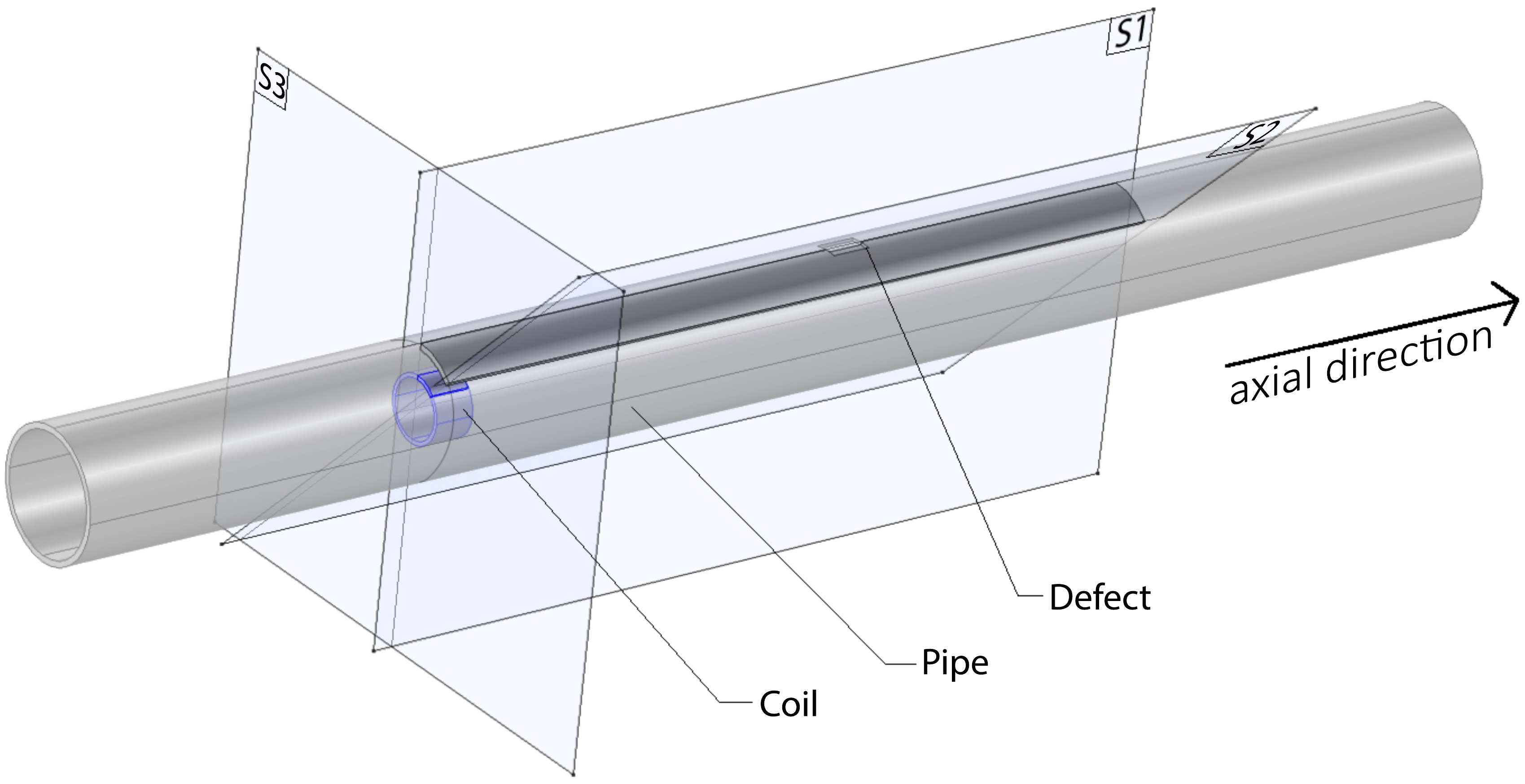}
	\caption{Illustration of the perfect pipe with a single Defect. The simplification of the simulation is done with the symmetries \textbf{S1}, \textbf{S2}, and \textbf{S3}.}
	\label{fig:simulation_geometry}
\end{figure}

The geometry used for this analysis uses the minimalist scenario of a perfect pipe with a single defect. To have a stable \gls{FEA} simulation, the mesh sizing has to be defined according to the wavelength of the electromagnetic field. It is standard practice to define the maximum size of a mesh element to be five times smaller than the wavelength~\cite{Marburg2008}. Following this rule, we define the maximum length of the mesh element $l_{max}$ according to the wavelength $\lambda$ of the electromagnetic field which is defined according to the material properties such as
\begin{equation}
\lambda = \dfrac{2\pi}{\sqrt{\dfrac{\omega\mu\sigma}{2}}},
\end{equation}
which $\mu$ the material permeability, $\sigma$ the material conductivity, and $\omega$ the wavelength frequency. $l_{max}$ the maximum length of the meshes in each medium is then defined as
\begin{equation}
l_{max}= \dfrac{\lambda}{5}.
\label{eq:element_length}
\end{equation}
This results in small mesh elements --- which are given in Table~\ref{tab:mesh_sizing} --- and a simulation which requires a large RAM and takes a lot of time to optimize. To simplify the simulation complexity, we use the symmetries referred as \textbf{S1}, \textbf{S2}, and \textbf{S3} on Figure~\ref{fig:simulation_geometry} resulting into a simulation which requires $90$Gb of memory and takes 40 minutes to compute on a 2x 3.1GHz Intel Xeon E5-2687W (8 Cores).

\begin{table}[b]
	\centering
	\newcolumntype{Y}{>{\centering\arraybackslash}X}
	\caption{Length of the maximum mesh element for a given material}
	\label{tab:mesh_sizing}
	\begin{tabularx}{0.3\textwidth}{Y Y}
		\hline Material & $l_{max}$ (m)\\ 
		\hline  Air & 0.1\\ 
		Soft-Iron & 0.012 \\ 
		Copper & 0.0115  \\ 
		\hline
	\end{tabularx}
\end{table}

Conversely to the axisymmetrical scenario, multiple measurements have to be taken in the remote area. In order to simulate the behavior of an array of receivers, we take an array of points measurement of both the magnetic field amplitude and the phase-shift along an array of virtual receivers. 

\begin{figure}[h]
	\centering
	\includegraphics[width=\linewidth]{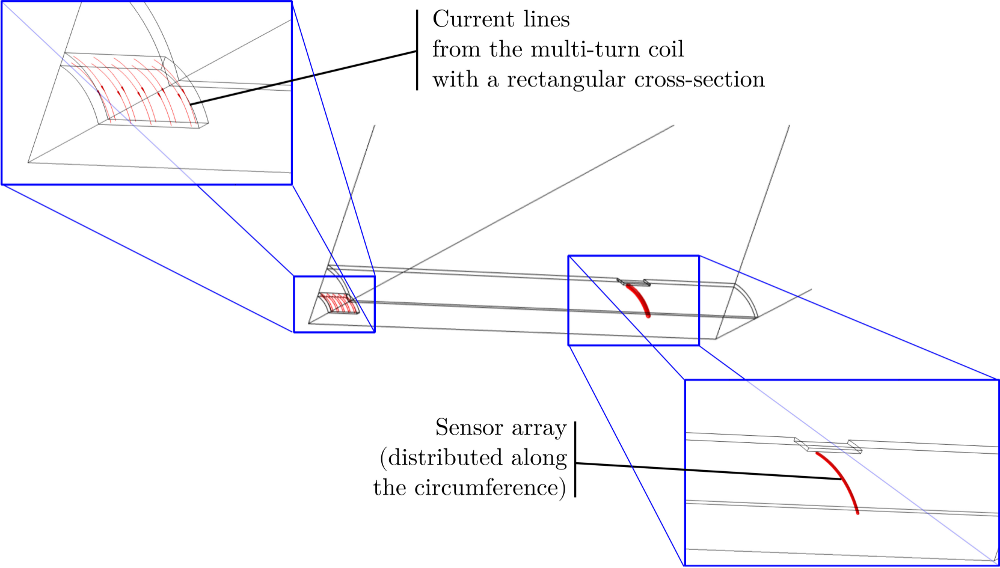}
	\caption{Display of the current lines in the exciter coil --- on the left --- and the array of virtual receivers --- on the right.}
	\label{fig:sensors_and_coil}
\end{figure}

\begin{figure}[h]
	\centering
	\subcaptionbox{\label{fig:amplitude}}{\includegraphics[width=0.45\linewidth]{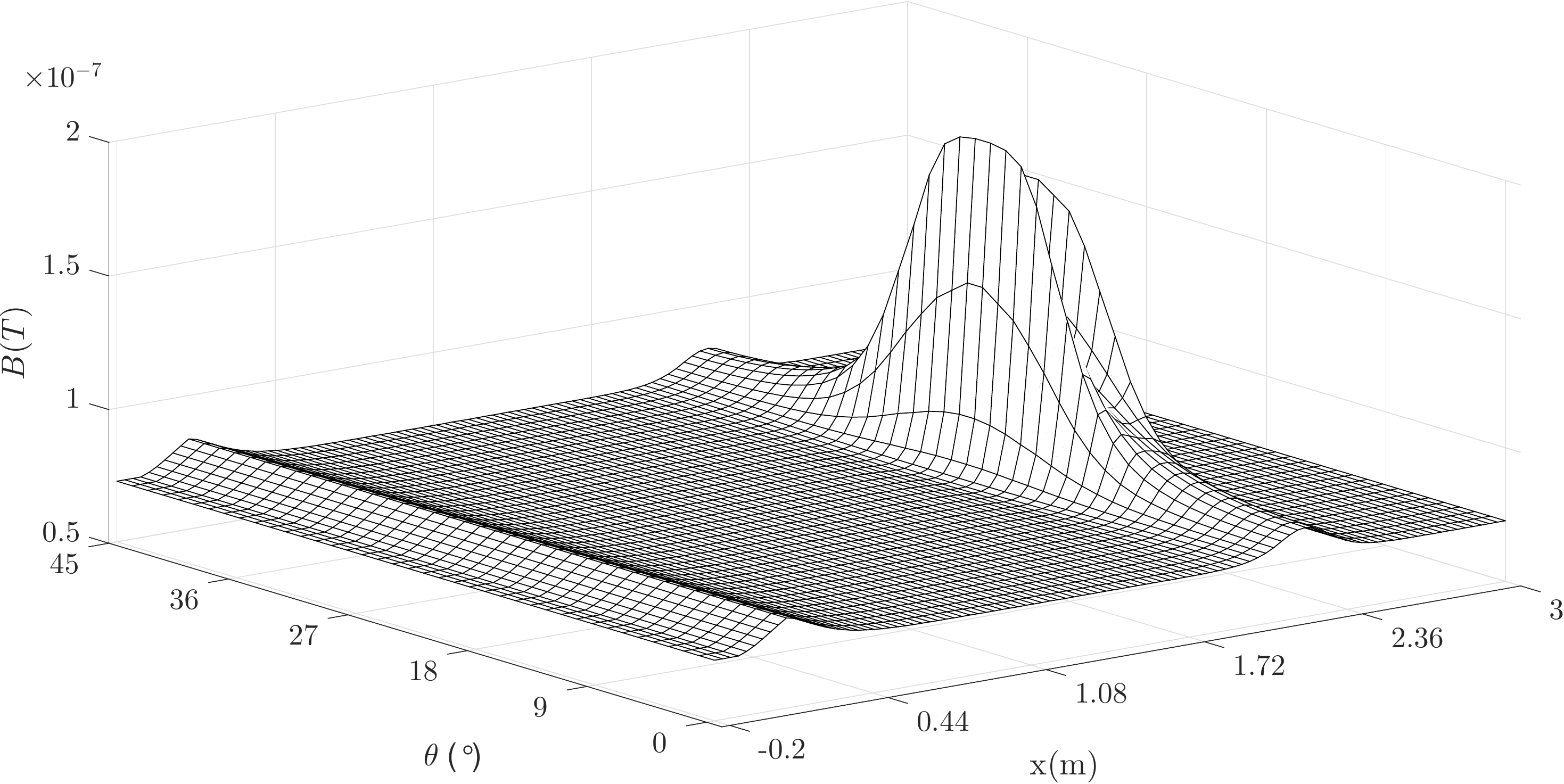}}
	\subcaptionbox{\label{fig:phaseSchift}}{\includegraphics[width=0.45\linewidth]{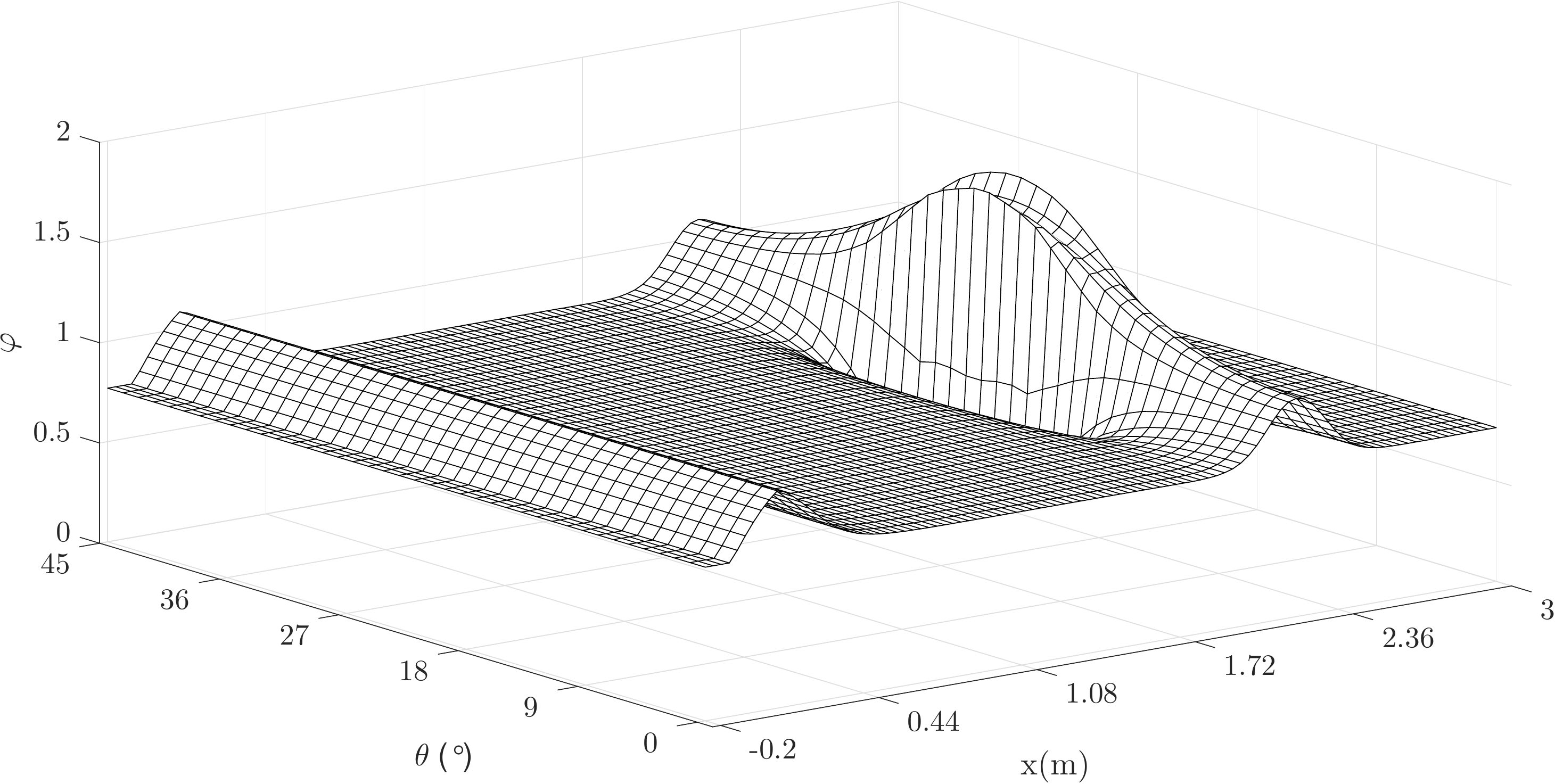}}
	\caption{Results of the 3D signal simulation. The amplitude of the magnetic field for the defect sweep is shown in~(\subref{fig:amplitude}), and the phase-shift is shown in~(\subref{fig:phaseSchift}).}
	\label{fig:result_3D_simulation}
\end{figure}

Using this \gls{FEA} configuration, we gradually change the position of the defect in the simulation in order to simulate the inspection of the pipe with an \gls{RFEC} tool. The measurement --- phase-shift and amplitude --- from the tool are mirrored over the symmetry plane \textbf{S1} and the measurements are shown in Figure~\ref{fig:result_3D_simulation}.

While more analysis has been done in \cite{Falque2014}, it can be seen from the simulation that there is a circumferential change of sensor measurements when the defect is above the exciter coil and a more local change in the sensor measurements when the defect is located above the sensor array. This phenomenon can be explained by the magnetic field getting homogenized while propagating outside of the pipe. We summarize these ideas in the following propositions:
\begin{prop}\label{pr:1}
	The measurement is attenuated by both the thickness located near the exciter coil and the receiver.
\end{prop}
\begin{prop}\label{pr:receiver_contribution}
	The attenuation due to the pipe thickness near the sensor array is reflected as a local offset in the gathered signal.
\end{prop}
\begin{prop}\label{pr:exciter_contribution}
	The attenuation due to the pipe thickness near the exciter coil is reflected as a circumferential offset in the gathered signal.
\end{prop}

\subsection{2D signal deconvolution}
\label{sec:2D_deconvolution}
Given the propositions defined in the preceding subsection, we can reformulate the Equation~\eqref{eq:inverse_pb_4} while accounting for the difference between an axisymmetrical tool and a tool equipped with an array of receivers:
\begin{equation}
\boldsymbol{Y}(x, \theta) = w_1 \boldsymbol{Y}^{\text{fg}}(x, \theta) + w_k \boldsymbol{y}^{\text{bg}}(x),
\label{eq:inverse_pb_5}
\end{equation}
where $\boldsymbol{Y}(x, \theta)$ is the sensor measurements from the array of receivers, $\boldsymbol{Y}^{\text{fg}}(x, \theta)$ is the deconvolved signal, and $\boldsymbol{y}^{\text{bg}}(x)$ is the background signal related to the attenuation of the electromagnetic field flowing through the pipe in the exciter area.

Solving the \gls{1D} signal deconvolution allows obtaining $\boldsymbol{y}^{\text{bg}}$. Therefore, the signal deconvolution is solved as a \gls{1D} problem such that
\begin{equation}
\overline{\boldsymbol{y}(x)} = w_1 \overline{\boldsymbol{y}^{fg}(x)} + w_k \boldsymbol{y}^{\text{bg}}(x),
\label{eq:inverse_pb_6}
\end{equation}
where $\overline{\boldsymbol{y}(x)}$ and $\overline{\boldsymbol{y}^{fg}(x)}$ are defined as
\begin{eqnarray}
\overline{\boldsymbol{y}(x)} \triangleq \dfrac{1}{2\pi}\int_{\theta = 0}^{2\pi} \boldsymbol{Y}(\theta, x)d\theta \label{eq:y_mean}\\
\overline{\boldsymbol{y}^{fg}(x)} \triangleq \dfrac{1}{2\pi}\int_{\theta = 0}^{2\pi} \boldsymbol{Y}^{fg}(\theta, x)d\theta.
\end{eqnarray}

While $\overline{\boldsymbol{y}(x)}$ can be computed, $\boldsymbol{Y}^{fg}$ is the deconvolved signal that we want to recover. In practice, $\overline{\boldsymbol{y}^{fg}(x)}$ is obtained simultaneously to $\boldsymbol{y}^{\text{bg}}$ through Algorithm~\ref{alg:de-convolution}, and $\boldsymbol{y}^{\text{bg}}$ is finally obtained by substituting the recovered $\boldsymbol{y}^{\text{bg}}$ into Equation~\eqref{eq:inverse_pb_5} as
\begin{equation}
\boldsymbol{Y}^{\text{fg}}(x, \theta) = \dfrac{1}{w_1} \big( \boldsymbol{Y}(x, \theta)- w_k \boldsymbol{y}^{\text{bg}}(x) \big).
\label{eq:inverse_pb_7}
\end{equation}
At this stage the inverse problem is not solved as the thickness of the pipe is not recovered. However, for some applications such as defect segmentation or localization, the signal deconvolution is sufficient.

\section{Results}
\label{sec:results}
The proposed algorithms are tested on two different datasets, with the first one based on simulated data and introduced in Sub-section~\ref{sec:mf_behavior}, and the second one based on a real dataset collected on a de-committed pipeline dedicated for research purpose.

\subsection{Simulated dataset}
We first discuss the performance of the signal deconvolution applied on the simulated data, which is used for validating that the algorithm is working properly.

\subsubsection{Data extrapolation}
We discuss here some technical aspects, which are required for making the signal deconvolution possible with our dataset. As described in Algorithm~\ref{alg:de-convolution}, having the gradient computed on a part of the signal --- the indices are from $i$ to $j$ --- leads to having the signal at the limit (i.e., $Y(x)|_{x = max(\boldsymbol{x})}$) not updated. With the spatial dependency of the signal deconvolution, this error drifts iteratively towards the center of the signal. This property of missing information is inherent to the \gls{RFEC} technology. We overcome this problem by padding the signal on each axial direction.

Furthermore, in Equation~\eqref{eq:y_mean}, $\overline{\boldsymbol{y}(x)}$ is calculated by integrating over $\theta$. However, the simulated data just covers the window $[0, \dfrac{\pi}{4}]$. We mirror the signal compared to the symmetry plane $\boldsymbol{S1}$, obtaining the signal for a simulation covering $[-\dfrac{\pi}{4}, \dfrac{\pi}{4}]$. In order to obtain the remaining part of the signal, we assume an absence of signal modification while the receivers, located at the angle $\pi$, are passing the defect; we then perform a B-spline interpolation for the remaining missing sensors.

\subsubsection{Signal deconvolution}
\begin{figure}[h]
	\centering
	\subcaptionbox{\label{fig:background}}{\includegraphics[width=0.45\linewidth]{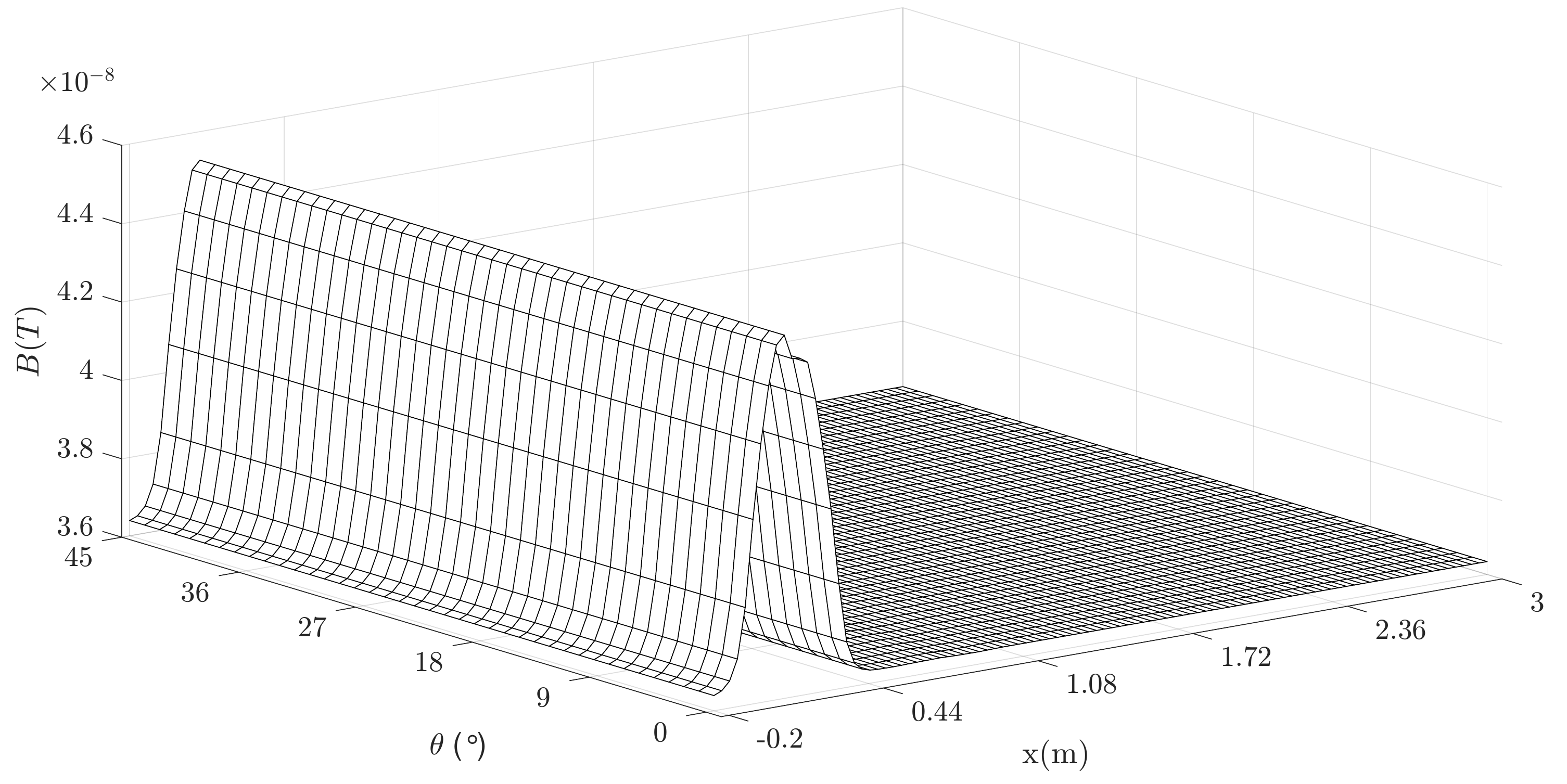}}
	\subcaptionbox{\label{fig:foreground}}{\includegraphics[width=0.45\linewidth]{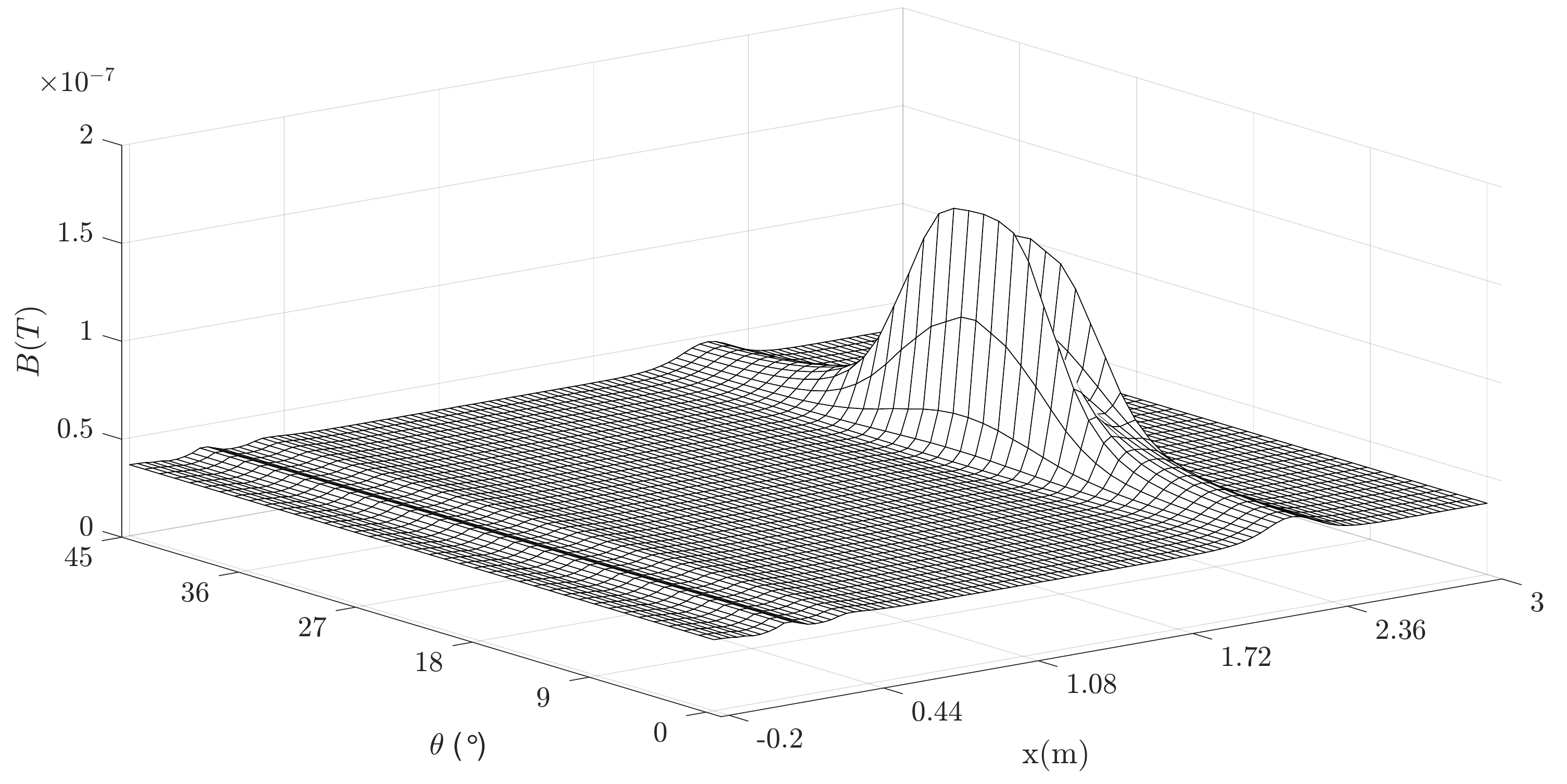}}
	\caption[Result of the 2D signal deconvolution on simulated data]{Result of the 2D signal deconvolution on simulated data. The estimated background~(\subref{fig:background}) is subtracted from $\boldsymbol{Y}$ to obtain the deconvolved signal~(\subref{fig:foreground}).}
	\label{fig:result_2D_deconvolution}
\end{figure}

The result of the signal deconvolution applied on the log-amplitude of the sensor measurement (i.e., $Y = log(B)$) is shown in Figures~\ref{fig:result_2D_deconvolution}(\subref{fig:background}) and \ref{fig:result_2D_deconvolution}(\subref{fig:foreground}). The background recovered --- on Figure~\ref{fig:result_2D_deconvolution}(\subref{fig:background}) --- is subtracted from the signal to obtain the deconvolved signal --- on Figure~\ref{fig:result_2D_deconvolution}(\subref{fig:foreground}).

Note that the scale between the Figure~\ref{fig:result_2D_deconvolution}(\subref{fig:foreground}) and \ref{fig:result_2D_deconvolution}(\subref{fig:background}) is different, which is expected from the generated signal shown in Figure~\ref{fig:result_3D_simulation}(\subref{fig:amplitude}).

\subsection{Real dataset}
\label{ap:real-dataset}
We now discuss the performance of the signal deconvolution applied on the real data.

\subsubsection{RFEC measurements gathering}
\label{sec:RFT}
\begin{figure}
	\centering
	\includegraphics[width=\linewidth]{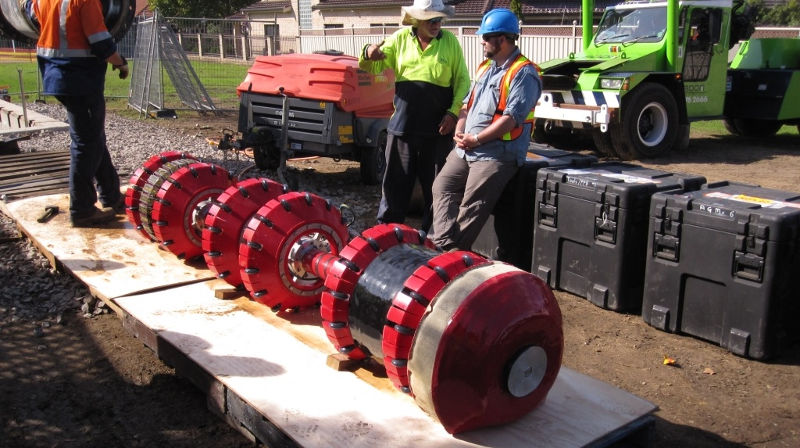}
	\caption{Commercial RFEC tool used for the field data collection in this work., courtesy of Russell NDE Systems Inc.}
	\label{fig:seeSnake}
\end{figure}

In the real dataset, the sensor measurements $\boldsymbol{Y}$ are obtained by scanning the entire pipeline using an inline commercial \gls{RFEC} tool. The design of the tool used for the data collection is similar to the one illustrated in Figure~\ref{fig:RFT_sensor_array}. In other words, the tool is composed of a single exciter coil, an array of electromagnetic receivers distributed along the circumferential direction, and a set of battery and electronic circuits used for on-board signal processing and gathering. A picture of the tool used  for the data collection is shown in Figure~\ref{fig:seeSnake}.

After inserting the RFEC tool inside the pipe through an insertion well\footnote{An insertion well and an extraction well have been created for this purpose prior to the inspection of the pipeline.}, the tool is pushed downstream in the pipeline by the water flow. While traveling down the pipe, the tool collects the sensor measurements the electromagnetic field amplitude and the phase-shift between the generated electromagnetic field and the gathered measurements --- as described in Section~\ref{sec:introduction} --- from the array of receivers. Simultaneously to the acquisition of the magnetic field measurements, the tool records information from several independent wheel encoders which, once merged together, provide an odometry to produce a rough estimation of the location.

\subsubsection{Laser scan collection}
\label{sec:LocalMap}

\begin{figure}
	\subcaptionbox{\label{fig:excavated-pipe}}{
		\centering
		\includegraphics[width=0.9\linewidth]{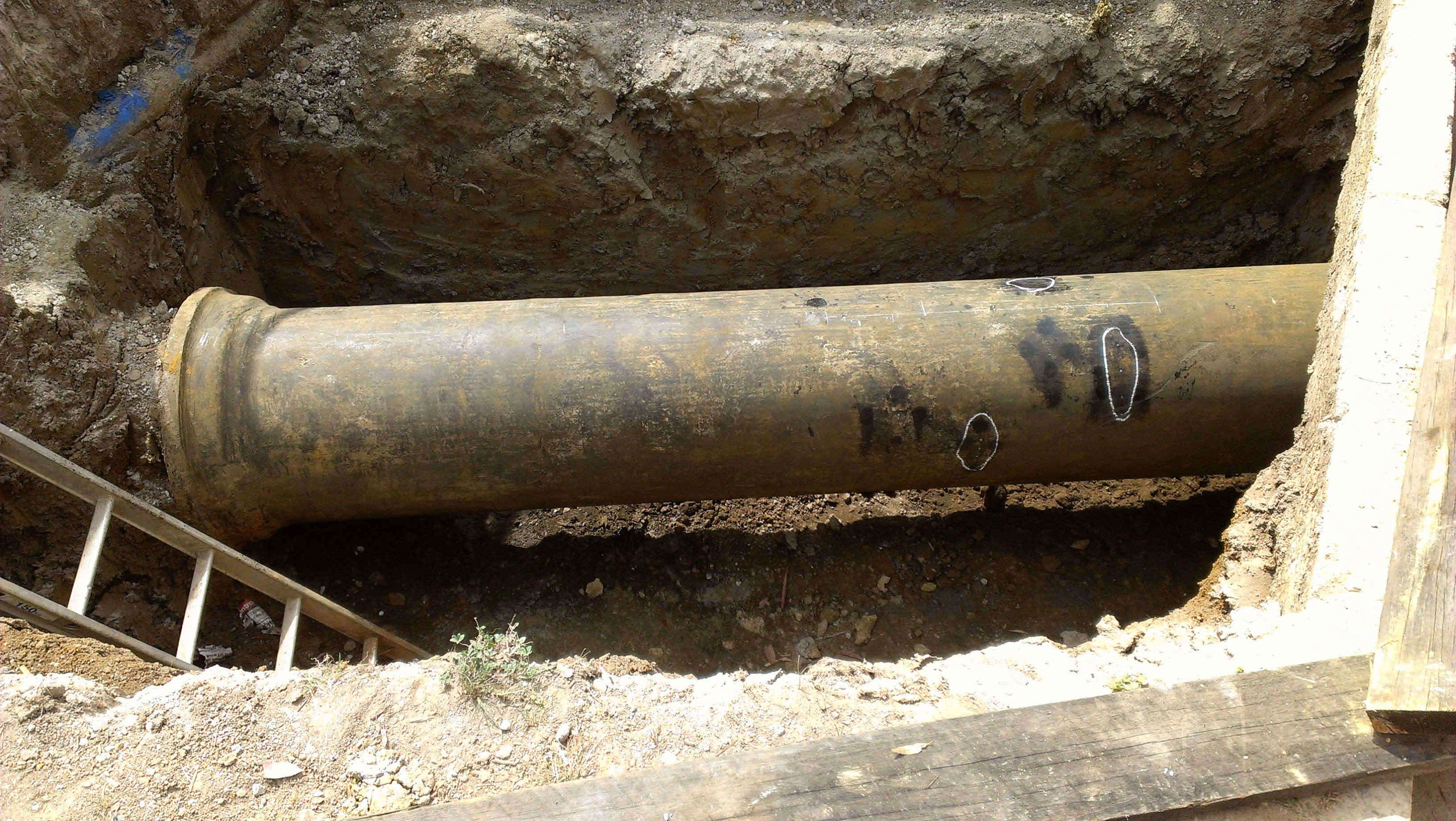}}
	\subcaptionbox{\label{fig:laser-scanner}}{
		\centering
		\includegraphics[width=0.9\linewidth]{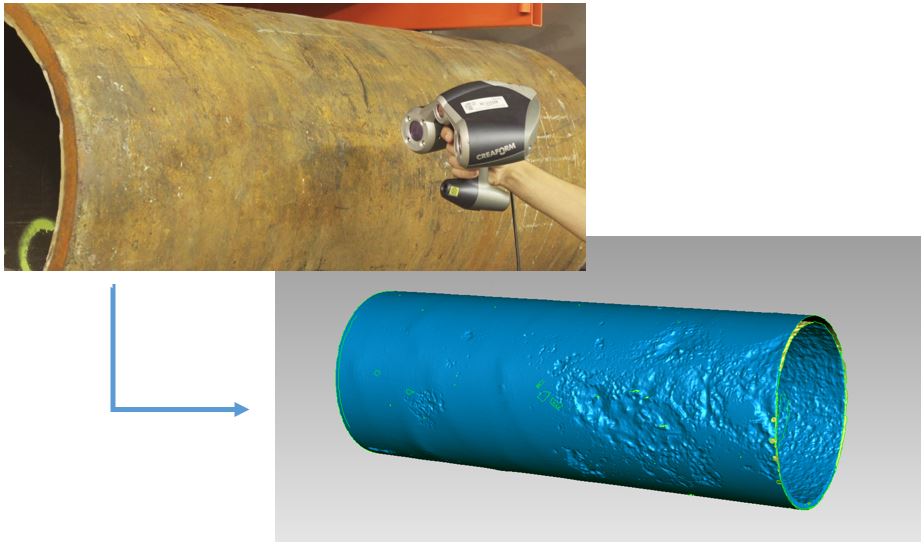}}
	\caption[processing of the raw data]{(\protect\subref{fig:excavated-pipe}) Pictures of an excavated pipe section with the Bell and Spigot joint visible on the left. (\protect\subref{fig:laser-scanner}) Digitalization of a 3D model of the pipe using a laser scanner. The 3D model is then transformed into a 2.5D thickness map.}
	\label{fig:ground_truth}
\end{figure}

To validate the performance of the proposed framework, a ground truth of the remaining non-corroded geometry of the pipe is required. To get such information, some of the pipe samples such as the one shown in Figure~\ref{fig:ground_truth}(\subref{fig:excavated-pipe}) are extracted and analyzed. Once excavated, pipe sections of about one meter long are cut and removed from the ground. A cleaning process, including manual cleaning and grit-blasting, is then started to remove corrosion. After cleaning, the remaining material consists of the non-corroded part of the pipe and can be considered as a reference for future analysis. However, the geometry has to be digitalized to be used for validation and in this case also as a local map.

The digitalization process is performed with a commercial high-resolution 3D laser scanner shown in Figure~\ref{fig:ground_truth}(\subref{fig:laser-scanner}). After scanning, the raw 3D model acquired by the laser scanner contains outliers and needs to be processed to be transformed into the same state space as the map $\boldsymbol{Y}$, which is a 2.5D thickness map. \gls{SOR} and \gls{ROR} filters~\cite{Rusu2011} are used in order to remove the outliers. The 3D model is then up-sampled into a uniform high dimensional point cloud using the \gls{RIMLS} algorithm~\cite{Oztireli2009}. After finding the cylindrical axis of the pipe segment, ray-casting is performed to find the thickness of the pipe. The rays are cast from the axis on regular intervals over $x$ and $\theta$. For each ray, if there is an intersection with two different voxels, then the Euclidean distance is allocated to the thickness map. At this stage of the process, the ground truth consists of a set of elevation maps $\boldsymbol{Y}^l$, which are the remaining non-corroded thickness of the pipe sections. More information about the methodology used for transforming the 3D model into a 2.5D thickness map is given in the work from~\cite{Skinner2014}.

The two \gls{RFEC} data and the laser scanned data are then aligned using a probabilistic framework that fuse all the information in order to find the most likely position of the laser-scanned pipe samples within the \gls{RFEC} sensor measurements of the pipeline~\cite{Falque2015}.

\subsubsection{Signal deconvolution}
\label{sec:generated-data_deconvolution-results}
Similarly to the simulated dataset, a padding is added to the initial sensor measurements prior to the signal deconvolution.

\begin{figure}
	\centering
	\subcaptionbox{\label{fig:gathered_dataset_depth} Laser-scans thickness of the inspected pipe thickness}
	{\includegraphics[width=\linewidth]{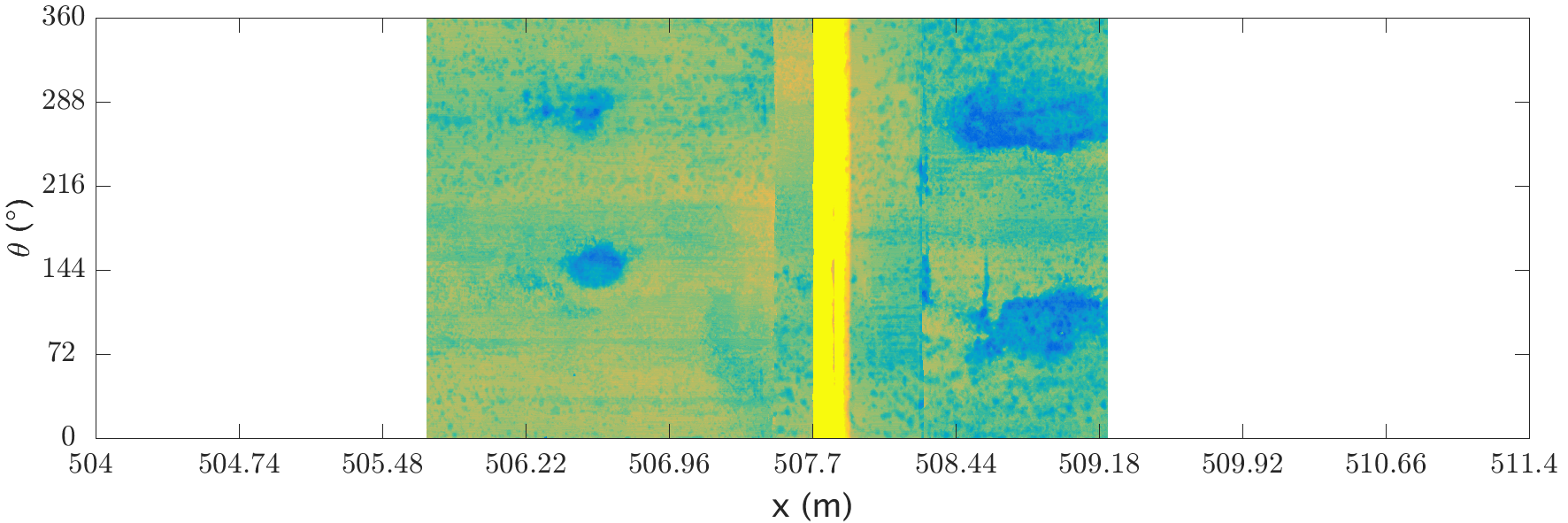}}
	\subcaptionbox{\label{fig:gathered_dataset_inverse_model} Corresponding sensor measurements}
	{\includegraphics[width=\linewidth]{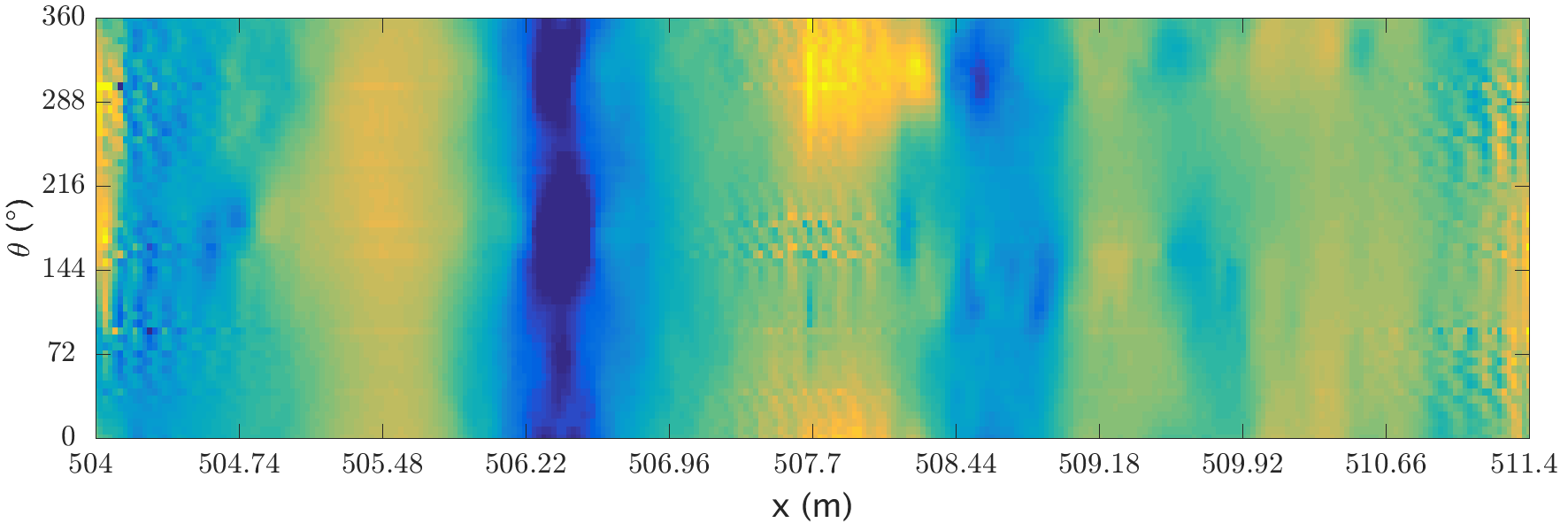}}
	\subcaptionbox{\label{fig:gathered_dataset_direct_model} Deconvolution algorithm applied on the sensor measurements}
	{\includegraphics[width=\linewidth]{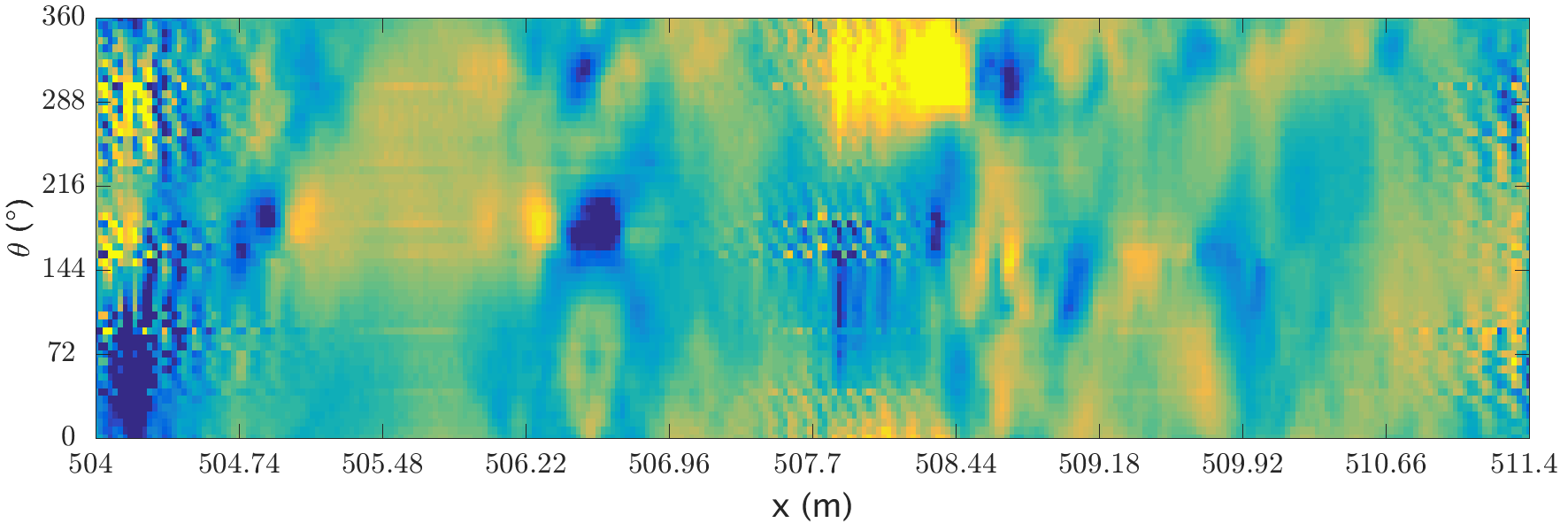}}
	\caption{Samples from the signal deconvolution applied to the gathered dataset. The pipe profile is shown in the first row, the associated RFEC signal is shown in the second row, and the applied signal deconvolution is shown in the last row. For both data set, darker colors relate to lower thicknesses --- the yellow stripes correspond to joints.}
	\label{fig:deconvolution_results}
\end{figure}

A sample of the deconvolution result is shown in Figure~\ref{fig:deconvolution_results}. The pipe profile is shown in \ref{fig:deconvolution_results}(\subref{fig:gathered_dataset_depth}), and the associated \gls{RFEC} data in \ref{fig:deconvolution_results}(\subref{fig:gathered_dataset_inverse_model}). The outcome of the signal deconvolution is shown in \ref{fig:deconvolution_results}(\subref{fig:gathered_dataset_direct_model}).

The magnetic field generated by the exciter coil flows through the path of least resistance. Therefore, the applied signal deconvolution --- which assumes the magnetic field to flow through a constant path --- is an approximation of the reality. In order to attenuate the effect of this approximation, a smoothing step is used in Algorithm~\ref{alg:de-convolution} (line $6$). In the real data sample, the filtering effect of the deconvolution is visible in \ref{fig:deconvolution_results}(\subref{fig:gathered_dataset_inverse_model}) where the circumferential offset at the axial position $506.5$m is removed after the deconvolution --- as shown in \ref{fig:deconvolution_results}(\subref{fig:gathered_dataset_direct_model}). As a result, the two large defects visible in the pipe profile \ref{fig:deconvolution_results}(\subref{fig:gathered_dataset_depth}) are highlighted after the signal deconvolution.

\section{Discussion}
\label{sec:discussion}
This journal provides a novel approach for performing the signal deconvolution on \gls{1D} and \gls{2D} sensor measurements gathered by tools designed similarly to the one shown in Figure~\ref{fig:RFEC_two_coil} and \ref{fig:RFT_sensor_array}. As discussed in Section~\ref{sec:method}, the signal deconvolution is a part of the inverse problem. We discuss in the following section the mapping from the sensor space to the pipe thickness space.

In order to map the deconvolved sensor measurements into the thickness space, a model has to be proposed for defining the function $g$. Considering the well known linear relationship between the deconvolved sensor measurements and the thickness space, a linear model such as $g(y) = ay +b$ would be sufficient. This linear relationship can be explained by the skin depth equation of a plane electromagnetic wave propagating through a ferromagnetic medium defined as follow:

\begin{equation}
\text{B}(t) =\underbrace{\text{B}_0 e^{-\sqrt{\dfrac{\omega \mu \sigma}{2}}t}}_\text{amplitude}
\underbrace{e^{-j\bigg(\sqrt{\dfrac{\omega \mu \sigma}{2}}t+\omega t\bigg)}}_\text{phase contribution},
\label{eq:mf-propagation}
\end{equation}
with B the magnetic field, $\text{B}_0$ the initial value of the magnetic field before penetrating the pipe material, $\omega$ the frequency, $\mu$ the magnetic permeability of the medium, $\sigma$ the electrical conductivity, and $t$ the distance traveled by the wave, i.e. the thickness of the pipe. The amplitude and the phase-lag are usually the measurements recorded by the \gls{RFEC} tools since they have a log-linear or linear relationship with the thickness of the conductive medium:
\begin{equation}
\left\{
\begin{array}{ll}
\phi_{\text{local}} = \sqrt{\dfrac{\omega \mu \sigma}{2}}t\\
ln(B)_{\text{local}} = ln(B_0) - \sqrt{\dfrac{\omega \mu \sigma}{2}}t
\end{array}
\right.
\label{eq:skin-depth-equation}
\end{equation}
While Equations~\eqref{eq:skin-depth-equation} are a simplification of the \gls{RFEC} phenomenon --- where the magnetic field flows from the exciter coil towards the receivers and the propagation through the ferromagnetic medium being just a portion of this path --- it is apparent that the relation between the sensor measurements and the thickness space is a function of the material properties. Thus a proper formulation for the linear model $g$ would be defined as $g(y) = a(\mu, \sigma)y+b(\mu, \sigma)$.

An interesting work in this direction has been published by Vasic et al. showing that the value of the material properties can be obtained from statistical analysis of the sensor measurements~\cite{Vasic2011}. Associating this work with the proposed signal deconvolution would be an interesting contribution and is considered as a future work.

The limitations of both models are similar to our prior work, in the sense that we assume the path of the magnetic field to be constant. Where in reality the magnetic field flows through the path of least resistance. Thus a linear model is an approximation of the reality and modeling the path of least resistance requires a non-linear model --- which would be much more complicated to train due to the much higher number of parameters required.

The aforementioned limitation is also one of the strengths of the proposed model. Indeed, while the training of a complex non-parametric model is feasible on simulated data, in practical scenario, the gathering of a real data set is extremely costly; therefore, there is a need for parametric models that can be trained on simulated data and later scaled to the real data set or models with low VC dimension that need a limited number of data for training.




\section*{Acknowledgment}
This publication is an outcome from the Critical Pipes Project funded by Sydney Water Corporation, Water Research Foundation of the USA, Melbourne Water, Water Corporation (WA), UK Water Industry Research Ltd,  South Australia Water Corporation, South East Water, Hunter Water Corporation, City West Water, Monash University, University of Technology Sydney and University of Newcastle. The research partners are Monash University (lead), University of Technology Sydney and University of Newcastle.

\bibliography{library}
\bibliographystyle{IEEEtranNoURL}

\end{document}